\documentclass[a4paper,11pt]{article}
\usepackage{jheppub}
\usepackage{graphicx}
\usepackage{amsmath}
\usepackage{amsfonts}
\usepackage{amssymb}
\def\thirdd{{\textstyle{\frac{1}{3}}}}
\def\OO{\mathcal{O}}

\def\Z{\mathcal{Z}}
\def\be{\begin{equation}}
\def\ee{\end{equation}}
\def\beqa{\begin{eqnarray}}
\def\eeqa{\end{eqnarray}}
\def\Eq#1{Eq.~\eqref{#1}}
\def\nn{\nonumber}

\usepackage{tensor}
\newcommand{\ARROW}[1]{\stackrel{\text{\makebox[0pt]{$#1$}}}{\longrightarrow}}
\newcommand{\tabref}[1]{Table~\ref{#1}}					
\newcommand{\figref}[1]{Figure~\ref{#1}}						

\newcommand{\tf}{\tau_{\text{F}}}
\newcommand{\tfl}[1]{\tau_{\text{F}_{#1}}}
\newcommand{\xt}{\tilde{x}}
\newcommand{\ttf}{\tilde{\tau}_{\text{F}}}
\newcommand{\xit}{\tilde{\xi}}
\newcommand{\dA}{d_{A}}

\title{Gradient-Flowed Thermal Correlators:
  How Much Flow is Too Much?}

\author{Alexander M.\ Eller, Guy D.\ Moore}

\affiliation{Institut f\"ur Kernphysik, Technische Universit\"at Darmstadt\\
Schlossgartenstra{\ss}e 2, D-64289 Darmstadt, Germany}
\emailAdd{meller@theorie.ikp.physik.tu-darmstadt.de,
guy.moore@physik.tu-darmstadt.de}

\abstract{
  Gradient flow has been proposed in the lattice community as a tool
  to reduce the sensitivity of operator correlation functions to noisy
  UV fluctuations.  We test perturbatively under what conditions doing
  so may contaminate the results.  To do so, we compute
  gradient-flowed electric field two-point correlators and stress
  tensor one- and two-point correlators at finite temperature in QCD.
  Gradient flow has almost no influence on the value of correlators
  until a (temperature- and separation-dependent) level of flow is
  reached, after which the correlator is rapidly compromised.  We
  provide a prescription for how much flow is ``safe.''
}

\begin{document}
\maketitle
\section{Introduction}
\label{sec:intro}

Gradient flow \cite{Narayanan:2006rf,Luscher:2009eq,Luscher:2010iy,%
  Luscher:2010we,Luscher:2011bx,Suzuki:2013gza,%
  Luscher:2013vga,Hieda:2016xpq,Suzuki:2016ytc}
is a nonperturbative and gauge-invariant method in quantum field
theory for defining not-quite-local operators with greatly improved
insensitivity to ultraviolet fluctuations.  Gradient flow is defined
by introducing a procedure which, configuration by configuration
within the Euclidean path integral, applies ``heat equation''
evolution to the fields, before constructing operators out of them.
Roughly speaking, one can think of this as replacing the fields in an
operator with those averaged over a Gaussian envelope.  However, by
using a nonlinear and gauge-invariant version of the heat equation,
the procedure maintains gauge invariance.  One can make a rigorous
connection between operators under gradient flow and renormalized
operators, and all perturbation theory tools needed to study
gradient-flowed operators have been developed
\cite{Luscher:2011bx}.

The main applications of gradient flow have been within lattice
quantum field theory.  The value of the $F^2$ operator ($F^{\mu\nu}$
the field strength) as a function of scale can be used to ``read off''
the scale-dependent coupling constant and therefore to perform scale
setting \cite{Luscher:2013vga}.
Gradient flow is also now widely used to remove UV fluctuations which
contaminate the determination of topology on the lattice
\cite{Berkowitz:2015aua,Borsanyi:2015cka,Petreczky:2016vrs,
  Taniguchi:2016tjc,Burger:2017xkz,Frison:2016vuc,Borsanyi:2016ksw}.
This is similar to older ``smearing'' methods
\cite{Berg:1981nw},
with the difference that the gradient-flow approach is on more solid
field theoretical foundations.

Gradient flow has also seen its first applications to the study of
thermodynamical properties of finite-temperature systems.
The energy density and pressure of SU(3) gauge theory were calculated directly on the lattice using gradient flow\cite{Asakawa:2013laa}. This was recently expanded to the energy--momentum tensor in order to determine the equation of state for SU(3) gauge theory \cite{Kitazawa:2016dsl}.
The great advantage of gradient flow in this context is that, by
reducing sensitivity to ultraviolet fluctuations, it can dramatically
reduce statistical fluctuations in evaluating thermal operator
expectation values and correlation functions.  For instance, consider
the determination of thermodynamical information.  One could evaluate
the energy density at temperature by evaluating the difference
$\langle T^{00} \rangle_{\beta} - \langle  T^{00} \rangle_{\mathrm{vac}}$,
the thermal-to-vacuum difference in the $00$ component of the stress
tensor.  The configuration-by-configuration squared fluctuations in
this quantity are set by the 2-point function
$\lim_{x\to 0} \langle T^{00}(x) T^{00}(0) \rangle$.  Based on
operator dimension, we see that this quantity diverges at small $x$ as
$x^{-8}$.  Of course on the lattice this divergence is cut off by the
lattice spacing and is $\OO(a^{-8})$.  This squared fluctuation must
be compared to $\beta^{-8}$, the square of the size of the energy
density difference; the number of spacetime points times
configurations must compensate this large ratio to obtain
a statistically significant measurement.  On the other hand, under
gradient flow to a depth $\tf$, we expect the overlapping 2-point
function to be $\OO(\tf^{-4})$ ($\tf$ has dimensions of length${}^2$,
not length).  Therefore the UV fluctuations which inhibit a
statistically significant evaluation are ameliorated and the number of
configurations we must evaluate to obtain good statistics is reduced
by a factor%
\footnote{%
  The factor is $(a^2/\tf)^2$ and not $(a^2/\tf)^4$ because there is
  only one independent measurement per $\tf^2$ of volume, rather than
  every $a^4$ of volume.}
of $(a^2/\tf)^2$.

A little gradient flow is certainly a good thing, improving
statistics, fixing some operator renormalization issues
\cite{Luscher:2010iy},
and making the lattice more continuum-like.  However, too much
gradient flow is definitely bad, as eventually we erase the
fluctuations responsible for the physics we want to study.  In
particular, we want to know, for the study of thermal one-operator and
multi-operator correlators, exactly how much gradient flow one may
apply before one changes the physics of interest.  In this note we
will study this problem perturbatively.  To our knowledge this is the
first perturbative study of thermal correlation functions, and of correlators of
spacetime-separated operators, under gradient flow.  Therefore we will
content ourselves for the moment with a leading-order perturbative
evaluation.  It is possible that interactions reveal some new physics
which makes the situation worse than what we find here, so it would be
valuable to extend these calculations to the loop level.  However we
will leave this for future work.

Here we will consider three types of correlation functions.  First and
simplest, we consider the stress tensor one-point function at finite
temperature.  As discussed above, this can be used to measure rather
directly the energy density as a function of temperature (if the
operator renormalization issues can be resolved; so far the
renormalization of a gradient-flowed stress tensor has only been
studied perturbatively
\cite{Luscher:2011bx,Hieda:2016xpq},
while a nonperturbative treatment is probably necessary).
Second, we will consider the correlator of two electric field
operators, embedded along a Polyakov line:
\begin{equation}
  \label{GEE}
  G^{^{\mathrm{EE}}}(\tau) = \frac{ \left\langle \mathrm{Re\:Tr\:}
    U(\beta,0;\tau,0) E_i(\tau,0) U(\tau,0;0,0) E_i(0,0)
    \right\rangle}
  {\left\langle \mathrm{Re\:Tr\:} U(\beta,0;0,0) \right\rangle}
\end{equation}
where $U(t_1,x_1;t_2,x_2)$ is a straight Wilson line from point
$(t_1,x_1)$ to point $(t_2,x_2)$ and $E_i$ is the electric field.
This operator was introduced in \cite{CaronHuot:2009uh},
who show that its analytical continuation to Minkowski frequency
determines the (momentum-space and coordinate-space) diffusion of a
heavy quark, $m_q \gg T$ in a thermal bath.  Recently there has been a
vigorous effort to measure this correlation function on the lattice
\cite{Francis:2015daa},
but so far only quenched results are available and the issue of the
$E$-field renormalization has not been resolved.  Gradient flow would
fix the renormalization issue and will hopefully improve statistical
power such that the correlation function can be reliably measured at
the nonperturbative level.  Finally, we will consider the correlation
function of two stress tensors at vanishing spatial momentum
(equivalently, integrated over spatial separation) as a function of
the Euclidean time separation $\tau$.  For those $T^{\mu\nu}$
components which couple to hydrodynamical modes, such as
$T^{00} T^{00}$ and $T^{0i} T^{0i}$, the correlator should be
$\tau$-independent and should reproduce thermodynamical information
(the heat conductivity and enthalpy density respectively).  For the
$\ell=2$ space component, \textsl{e.g.}\ $T^{xy} T^{xy}$, the
analytical continuation of the correlator holds information about the
shear viscosity
\cite{Zubarev,Karsch:1986cq,Meyer:2007ic,Meyer:2011gj}.

In the next section we will develop perturbative tools for gradient
flow at finite temperature in coordinate space, which turns out to be
the most convenient for the problems we study here.  Next, Section
\ref{sec:calc} contains the specific details of the leading-order
calculations of each correlator mentioned above.  In every case we
find that there is a ($\tau$-dependent) range of flow times $\tf$ for
which the correlator feels exponentially suppressed corrections; but
for more flow it quickly goes wrong.  We end with a discussion which
presents our recommendations for the amount of flow which can be
applied ``safely,'' given our lowest-order perturbative results.

\section{Gradient flow at temperature in coordinate space}
\label{sec:tec}

We write the unflowed gauge field as $A_\mu^a(x)$ and will generally
suppress the color index $a$.  The flowed gauge field $B_\mu(x,\tf)$ is
defined at nonnegative flow time $\tf$ through the $\tf=0$ boundary
condition
\be
\label{eq:InitialCondition}
\left.\tensor{B}{_\mu}(x,\tf) \right|_{\tf=0}=\tensor{A}{_\mu}(x) 
\ee
and the flow equation
\be
\label{eq:flowequationSU(N)}
\frac{\partial \tensor{B}{_\mu}(x,\tf)}{\partial\tf}=\tensor{D}{_\nu} \tensor{G}{_\nu_\mu}(x,\tf) + \alpha_{0} \tensor{D}{_\mu} \tensor{\partial}{_\nu} \tensor{B}{_\nu}(x,\tf),
\ee
where $\tensor{G}{_\nu_\mu}(x,\tf)$ is the field strength
tensor written using $B_\mu(x,\tf)$ rather than $A_\mu(x)$.
The second term in the flow equation constitutes as $\tf$-dependent
gauge choice, which is convenient to make in the context of
perturbative calculations \cite{Luscher:2010iy}.
Choosing $\alpha_0=1$ and working to linearized order, the flow
equation simplifies to
\be
\label{eq:flowfree}
\frac{\partial \tensor{B}{_\mu}(x,\tf)}{\partial \tf} = \tensor{\partial}{_\nu}  \tensor{\partial}{_\nu} \tensor{B}{_\mu}(x,\tf), 
\ee
which is the heat equation. 

In vacuum, the Feynman-gauge momentum-space propagator after flow is
\be
\label{eq:BBCorr}
\tensor*{G}{^B^B_{\text{E}}}(p,\tfl1,\tfl2) = \int d^4 x e^{ip_\mu x^\mu}
\left\langle \tensor*{B}{_\mu}(x,\tfl1) \tensor*{B}{_\nu}(0,\tfl2) \right\rangle
= g^{2}  \tensor{\delta}{_\mu_\nu}  \frac{e^{-(\tfl1 + \tfl2) p^2}}{p^2} 
\ee
and Fourier transforming leads to the zero temperature flowed
propagator in coordinate space
\be
\label{eq:BBCorrX}
\tensor*{G}{^B^B_{\text{E}}}(x,\tfl1,\tfl2) =
\frac{ g^{2} \tensor{\delta}{_\mu_\nu}}
     {4 \pi^2 x^{2}}\left(1-e^{- \frac{x^{2}}{4 (\tfl1 + \tfl2)}} \right).
\ee
This result was found by L\"uscher in \cite{Luscher:2010iy}.
But we could have reached this result faster by noting that the
coordinate-space propagator before flow is the solution to the Poisson
equation $-\partial_\mu^2 G^{AA}_\text{E}(x) = g^2 \delta^4(x)$, which is
$G^{AA}_\text{E}(x) = g^2/(4\pi^2 x^2)$.  At tree level and in Feynman
gauge, flow is the application of the heat equation to this
propagator, which is the same as convolving it with a Gaussian
envelope,
\be
\label{eq:convolve}
G^{BB}(x,\tfl1,\tfl2)= \int d^{4}y \frac{g^2}{4 \pi^{2} y^{2}}
  e^{-\frac{(x-y)^2}{4(\tfl1 + \tfl2)}}= \frac{g^2}{4 \pi^{2} x^{2}}
  \left(1-e^{- \frac{x^{2}}{4 (\tfl1 + \tfl2)}} \right).
\ee
Alternatively, one may take the right-hand expression as an
\textsl{Ansatz} and verify that it satisfies the $\tf=0$ boundary
conditions and the heat equation.

To introduce finite temperature, we restrict the Euclidean time to lie
in $x^0 \in [0,\beta]$ with periodic boundary conditions.  The fast
way to find the coordinate-space propagator is to note that the
Poisson equation is now solved using the method of images;
\be
\label{eq:images}
G^{AA}_{\text{E},\beta}(x^0,\vec{x}) = \sum_{n\in \Z}
\frac{g^2}{4\pi^2 x_n^2} \,, \qquad
x_n^\mu \equiv (\, x^0{+}n\beta \, , \, \vec{x} \,) \,,
\ee
and that flow again corresponds to evolving this propagator under the
heat equation or convolving with a Gaussian:
\be
\label{eq:BBCorrXT}
\tensor*{G}{^B^B_{\text{E},\beta}}(x,\tfl1,\tfl2)=
\sum_{n= - \infty}^{\infty}  \frac{ g^{2} \tensor{\delta}{_\mu_\nu}}
    {4 \pi^2 x_n^2}
    \left(1-e^{- \frac{x_n^2}{4(\tfl1 + \tfl2)}} \right).
\ee 
We could also arrive at this result the ``hard way'' by Fourier
transforming the finite-temperature, flowed momentum-space propagator
\be
\label{eq:BBPT}
\tensor*{G}{^B^B_{\text{E}}}(x,\tfl1,\tfl2)=
  T \sum_{p^{0}=2\pi mT}^{m\in \Z}
  \int \frac{d^{3}p}{(2 \pi)^{3}} e^{i p \cdot x}
 \times \frac{g^2 e^{-p^2(\tfl1+\tfl2)}}{p^2}
\ee
by use of Poisson's summation formula \cite{stein1971introduction}
\be
\label{eq:Poisson}
T \sum_{p^{0}=2\pi mT}^{m\in \Z} = \sum_{n\in \Z}
\int_{- \infty}^{\infty} \frac{d p^{0}}{2 \pi} e^{i p^{0} \beta n}
\ee   
to rewrite the summation over $p^{0}$ as a sum over coordinate-space
copies -- essentially, the same images as above.  At this point each
$\int dp^0$ term
represents a vacuum contribution with a different $x^0$ position,
shifted into one of the image copies.  This leads rather directly back
to \Eq{eq:BBCorrXT}.  In the following we will only work at finite
temperature so we will suppress the subscript $\beta$.

\begin{figure}[tb]
  \hfill \includegraphics[width = 0.5\textwidth]{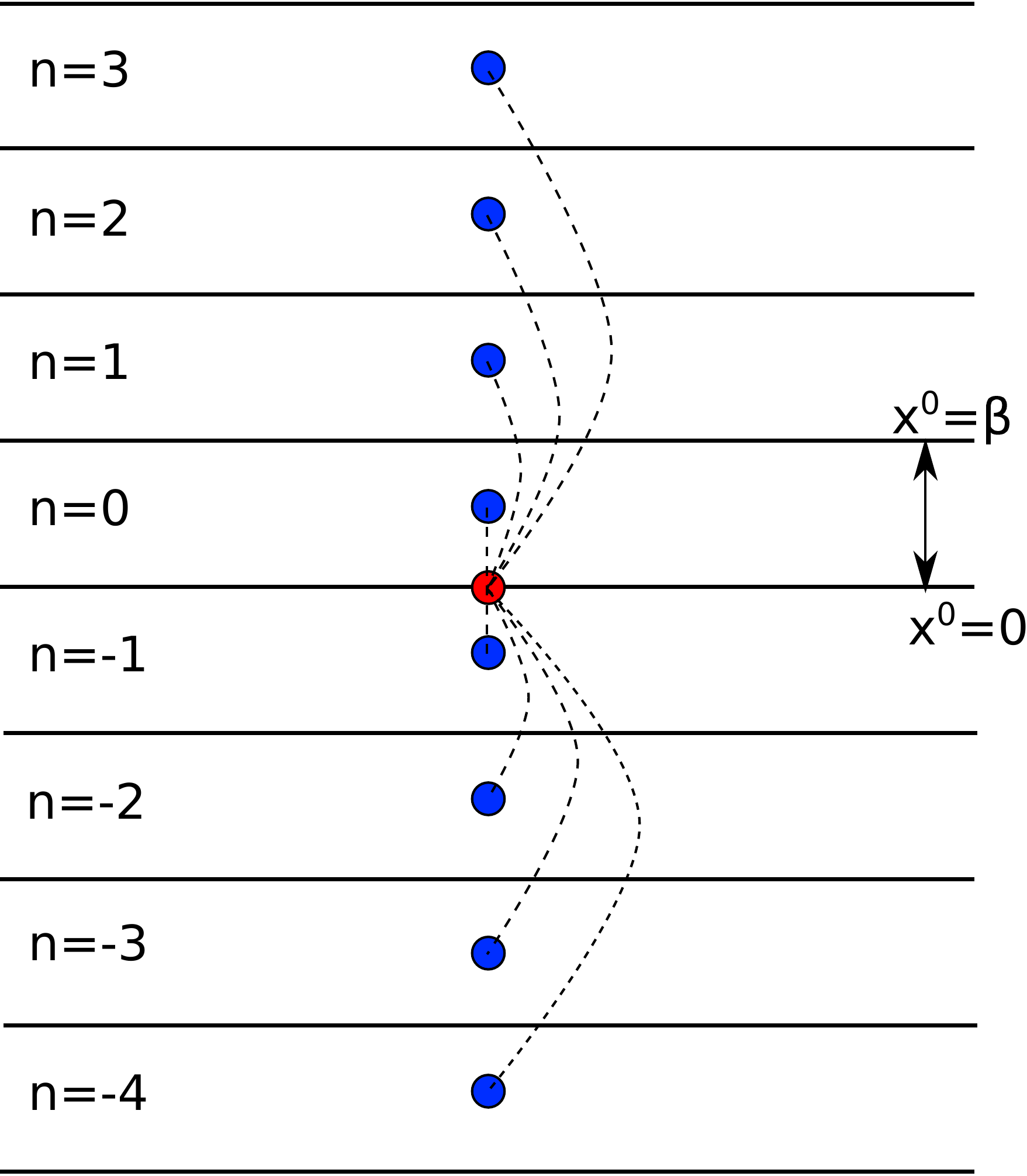}
  \hfill $\phantom{.}$
  \caption{\label{fig:ImageCartoon}
   Cartoon of the images and correlation of the finite--temperature
   gauge--field propagator. The colored circles represent the
   gauge fields, the horizontal lines the periodic boundaries in the
   time plane, and the dashed lines the correlations between one field
   and the images of the other.}
\end{figure}

When we take correlation functions, we will have to include a sum over
images for \textsl{each} propagator which appears.
We present a cartoon of this procedure in \figref{fig:ImageCartoon}.

\section{Calculations}
\label{sec:calc}

We will now use the coordinate-space propagator to compute the desired
correlation functions.  While the last section has introduced
propagators between fields with different amounts of flow, here we
will only consider correlators where all operators are evaluated after
the same amount of flow $\tf$; so the previous formulae should be
modified by writing $\tfl1=\tfl2=\tf$.

\subsection{Energy-Momentum Tensor One--Point Function}
\label{subsec:T}

The tree-level energy--momentum tensor in Yang--Mills theory is%
\footnote{At the loop level we would need to include the trace anomaly.}
\begin{eqnarray}
\label{eq: Def energy-momentum te}
\tensor*{T}{^B_\mu_\nu}(x,\tf)&=& \frac{1}{g^{2}}\left[\left(
  \tensor*{G}{_\mu_\sigma^a} \tensor*{G}{_\nu_\sigma^a} \right)
  (x,\tf) -\frac{1}{4} \tensor{\delta}{_\mu_\nu} \left(
  \tensor*{G}{_\omega_\sigma^a} \tensor*{G}{_\omega_\sigma^a}
  \right)(x,\tf)\right],
\\
G_{\mu\nu}(x,\tf) & = & \partial^{x}_\mu B_\nu(x,\tf)
- \partial^{x}_\nu B_\mu(x,\tf) \nn \,.
\end{eqnarray}
We evaluate the correlator of two field strengths by splitting the
field strengths to reside at points $x,y$, write
\be
\langle B_\mu(x,\tf) B_\nu(y,\tf) \rangle = G^{BB}_{\text{E}}(x-y,\tf)
\ee
which we found in \Eq{eq:BBCorrXT}, take derivatives, and then set
$x=y$.  Introducing a dimensionless rescaled flow time
$\ttf \equiv 8\tf / \beta^2$, we evaluate the two field strength
correlators which we need,
\beqa
\left\langle \tensor*{G}{_0_\sigma^a} \tensor*{G}{_0_\sigma^a}
\right\rangle &=& \frac{3 g^{2} \dA}{\pi^{2} \beta^{4}}
\sum_{n\in \Z} \left[ e^{-\frac{n^{2}}{\ttf}}\cdot \left( \frac{1}{\ttf^{2}} +  \frac{1}{\ttf} \frac{1}{n^{2}} +\frac{1}{n^{4}} \right) - \frac{1}{n^{4}} \right],
\\
\left\langle \tensor*{G}{_i_\sigma^a} \tensor{G}{_i_\sigma^a}
\right\rangle &=& \frac{3 g^{2} \dA}{\pi^{2} \beta^{4}}
\sum_{n\in \Z} \left[ e^{-\frac{n^{2}}{\ttf}}\cdot \left( \frac{1}{\ttf^{2}} -  \frac{1}{\ttf} \frac{1}{n^{2}} - \frac{1}{n^{4}} \right) + \frac{1}{n^{4}} \right],
\eeqa
where each $n^2$ arises as $x_n^2/\beta^2$.  Then we
combine them to find a closed expression for the stress-tensor
one-point function after flow,
\be
\label{eq: T one point sum}
\left\langle \tensor{T}{_0_0} \right\rangle = - \left\langle
\tensor{T}{_i_i} \right\rangle = \frac{3 \dA}{\pi^{2}
  \beta^{4}} \sum_{n \in \Z} \left[ e^{-\frac{n^{2}}{\ttf}}\cdot \left( \frac{1}{2~\ttf^{2}} +  \frac{1}{\ttf} \frac{1}{n^{2}} +\frac{1}{n^{4}} \right) - \frac{1}{n^{4}} \right].
\ee
Here $\dA=N_c^2-1=8$ is the dimension of the group, which counts gluon
colors.  The sum over $n$ is a sum over images; the vacuum result is
the $n=0$ term, which is defined as the $n\to 0$ limit and which
actually vanishes.  In the $\tf\to 0$ limit the exponential terms vanish
and we have only the $1/n^4$ term, confirming as expected that
\be
\left\langle \tensor{T}{_0_0} \right\rangle = -\frac{3
  \dA}{\pi^{2} \beta^{4}} \sum_{n \neq 0} \frac{1}{n^{4}} =  -\frac{6 \dA}{\pi^{2} \beta^{4}} \zeta(4),
\ee
the standard Stefan-Boltzmann result.

In the opposite limit, $\ttf \gg 1$, many terms contribute to the sum
and we may approximate it with an integral, giving rise to
\beqa
\langle T_{00} \rangle_\beta  & \underset{\ttf \gg 1}{\longrightarrow}
& \frac{3 \dA}{\pi^{2} \beta^{4}}
\frac{1}{\ttf^{3/2}} \int_{-\infty}^{\infty}~du \frac{1}{u^{4}}
\left[ -1 + \left( 1+ u^{2} +\frac{u^{4}}{2}\right) e^{-u^{2}} \right]
\nn  \\
&=& - \frac{3 \dA}{\pi^{2} \beta^{4}} \frac{1}{\ttf^{3/2}} \frac{\sqrt{\pi}}{6}
= - \frac{\dA}{32 \sqrt{2} \pi^{3/2} \tf^{3/2} \beta} .
\eeqa
This result corresponds to the contribution arising from the zero
Matsubara frequency, as all other Matsubara frequencies are damped
away by the flow.

\begin{figure}[tb]
  \hfill \includegraphics[width = 0.8\textwidth]{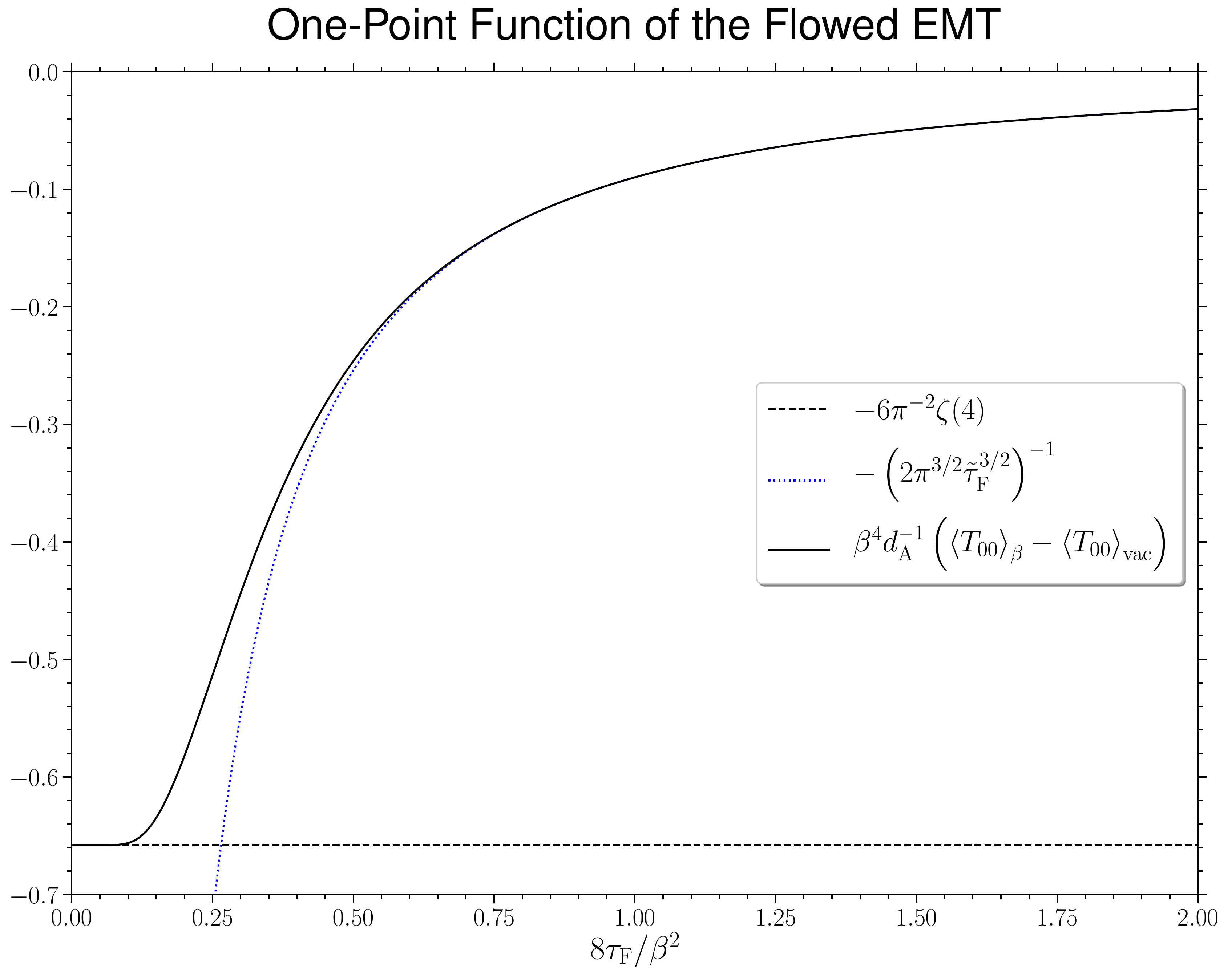}
  \hfill $\phantom{.}$
  \caption{\label{fig: T one point Sum Plot}
    Plot of the one--point function of the energy--momentum tensor as
    a function of the applied gradient flow, together with its
    asymptotic small $\tf$ and large $\tf$ behavior.} 
\end{figure}

For finite $\ttf$ we evaluate the sum numerically and display the
result in \figref{fig: T one point Sum Plot}.  The plot shows that,
for small $\ttf$, the corrections to Stefan-Boltzmann are
exponentially small, physically representing the exponentially small
amplitude for the ``smearing'' due to flow to stretch all the way
around the periodic direction.  However the stability of the result
then rather abruptly breaks down above $\ttf \sim 0.12$, and for large
$\ttf$ values the thermal contribution is almost completely lost.  If
we require that the flow change the determined energy density by at
most $1\%$, then we can constrain the allowed flow depth to be
$8 \tf / \beta^{2} \leq 0.12$.  On the lattice with $N_t$ lattice
points around the temporal direction, that corresponds to
$\tf/a^2 \leq 0.015 N_t^2$ with $a$ the lattice spacing.

\subsection{Electric--Field Correlation Function at Finite Temperature}
\label{subsec:EE}
In \Eq{GEE} we see that the electric field correlator of interest
contains Wilson lines forming a Polyakov loop.  However in a
lowest-order evaluation these are irrelevant, and only derivatives of
the gauge-field propagator are involved.  The leading-order
contribution reads
\be
\left\langle\tensor*{E}{_i^a}(x,\tf)
  \tensor*{E}{_j^b}(0,\tf)   \right\rangle =
  \left. \tensor*{\partial}{_0^x} \tensor*{\partial}{_0^y}
  \left\langle \tensor*{B}{_i^a}(x,\tf)
  \tensor*{B}{_j^b}(y,\tf)\right\rangle
  + \tensor*{\partial}{_i^x} \tensor*{\partial}{_j^y}
  \left\langle \tensor*{B}{_0^a}(x,\tf)
  \tensor*{B}{_0^b}(y,\tf)\right\rangle \right|_{y=0}.
\ee
Differentiating and introducing the dimensionless scaled coordinate
$\xt_n = x_n/\beta$ and the ratio of squared coordinate to flow time
$\xit_n^2 = \xt_n^2/\ttf$, we find
\beqa
\nn
  \left\langle\tensor*{E}{_i^a}(x,\tf)
  \tensor*{E}{_j^b}(0,\tf) \right\rangle
 & = & \frac{g^{2} \tensor{\delta}{^a^b}}{\pi^{2}}
  \sum_{n\in \Z} \frac{1}{\xt_n^{4}}
  \left[ \frac{\delta_{ij}(\xt_n^{0})^{2}+\xt_{i}\xt_{j}}{\xt_n^{2}}
    \left(( \xit_n^{4}{+}2 \xit_n^{2}{+}2)  e^{-\xit_n^{2}} - 2 \right) \right.
\\
& & \hspace{2.4cm} {} + \left. \delta_{ij}
\left( 1 - (1{+}\xit_n^{2}) e^{-\xit_n^{2}}\right)
 \vphantom{\frac{\delta_{ij}}{\xt_n^2}}  \right].
\eeqa
In this expression we have allowed the electric fields to be at
different spatial coordinates, but the correlator relevant for heavy
quark transport involves $\vec{x}=0$, which we will set from now on.
Our result then simplifies to
\be
\label{eq: EE Temp Sum}
\left\langle\tensor*{E}{_i^a}(x^{0},\tf)
\tensor*{E}{_j^b}(0,\tf) \right\rangle =
\frac{g^{2} \tensor{\delta}{^a^b}}{\pi^{2} \beta^4}
\sum_{n\in \mathcal Z}
\frac{\tensor{\delta}{_i_j}}{\xt_n^{4}}
\left[ ( \xit_n^{4} +  \xit_n^{2} + 1 )  e^{-\xit_n^{2}} -1  \right].
\ee
This is the main result of this section.

To explore this result further, we consider first the limit of small
flow time, $\ttf\to 0$ or $\xit \to \infty$.  In this limit
$\left( \xit^{4} +  \xit^{2} + 1 \right)  e^{-\xit^{2}} \simeq 0$.
The sum can be performed analytically and the result is 
\be
\left\langle\tensor*{E}{_i^a}(x,\tf) \tensor*{E}{_j^b}(0,\tf) \right\rangle=- \frac{\pi^{2} g^{2} \tensor{\delta}{^a^b} \tensor{\delta}{_i_j} }{\beta^{4}} \frac{\cos(2 \pi \xt^{0})+ 2}{3\sin^{4}(\pi \xt^{0})}.
\ee
The correlation function is negative, as expected; the electric field
is odd under the time-reflection operator
\be
E \ARROW{\Theta} -E\,,
\ee
and so its correlation function should be negative.  Note however that
the time-integrated $\langle EE \rangle$ correlator could still be
positive due to contact terms when the operators overlap.

We can also explore the opposite limit of large flow time,
$\ttf \gg 1$, which allows us to approximate the sum over $n$ with an
integral,
\beqa
  \left\langle\tensor*{E}{_i^a}(x,\tf) \tensor*{E}{_j^b}(0,\tf)
  \right\rangle
   & \underset{\ttf \gg 1}{\longrightarrow} &
  \frac{g^{2} \tensor{\delta}{^a^b}}{\pi^{2}\beta^4}
  \int_{-\infty}^{\infty}~dn \: \frac{\delta_{ij}}{\xt_n^4}
  \left[ ( \xit_n^{4} +  \xit_n^{2} + 1 )  e^{-\xit_n^{2}} -1  \right] 
\nn \\
&=&  \frac{g^{2} \tensor{\delta}{^a^b}}{\pi^{2}}
\frac{\tensor{\delta}{_i_j}}{\beta^{4} \ttf^{3/2}}
\int_{-\infty}^{\infty}~du \frac{1}{u^{4}} \left[ -1 + \left( 1 +
  u^{2} + u^{4} \right) e^{-u^{2}} \right] 
\nn \\
&=& \frac{g^{2} \tensor{\delta}{^a^b} \tensor{\delta}{_i_j}}{48 \sqrt{2} \pi^{3/2} \tf^{3/2} \beta },
\eeqa
which is the same result we would get by considering only the
contribution of the zero Matsubara frequency.  In contrast to the
small $\ttf$ limit, this result is positive.  There is no
contradiction with fundamental theorems, because the operator after
flow is no longer local, so $\Theta$-odd behavior does not
ensure negative correlations.  But this indicates that the result at
large flow times has been thoroughly contaminated with contact-term
type contributions.  Once a correlator which is expected to be
negative becomes positive due to flow, the character of the
correlation function has been fundamentally altered.

\begin{figure}[tb]
 \includegraphics[width = 0.48\textwidth]{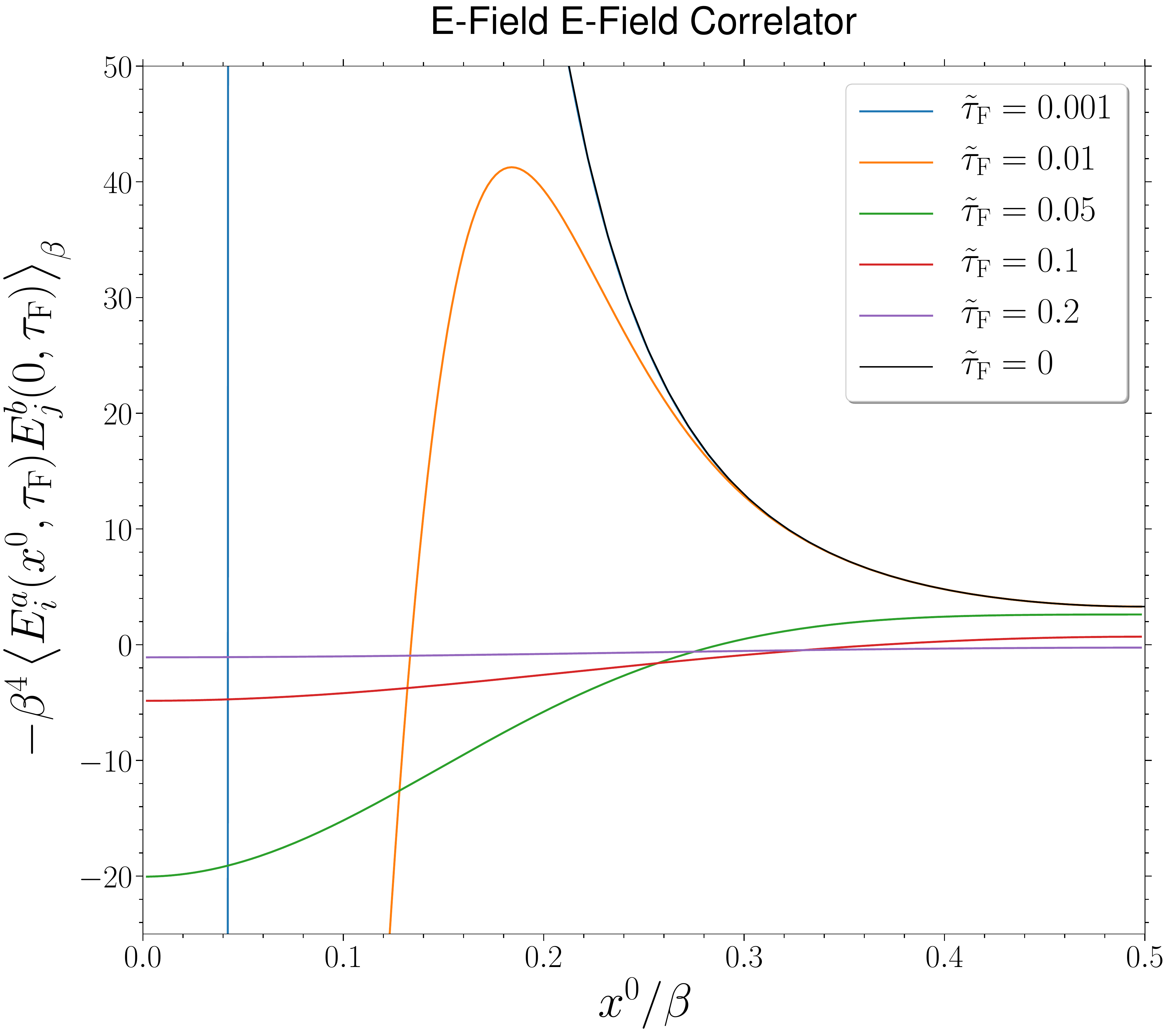} \hfill
 \includegraphics[width = 0.5\textwidth]{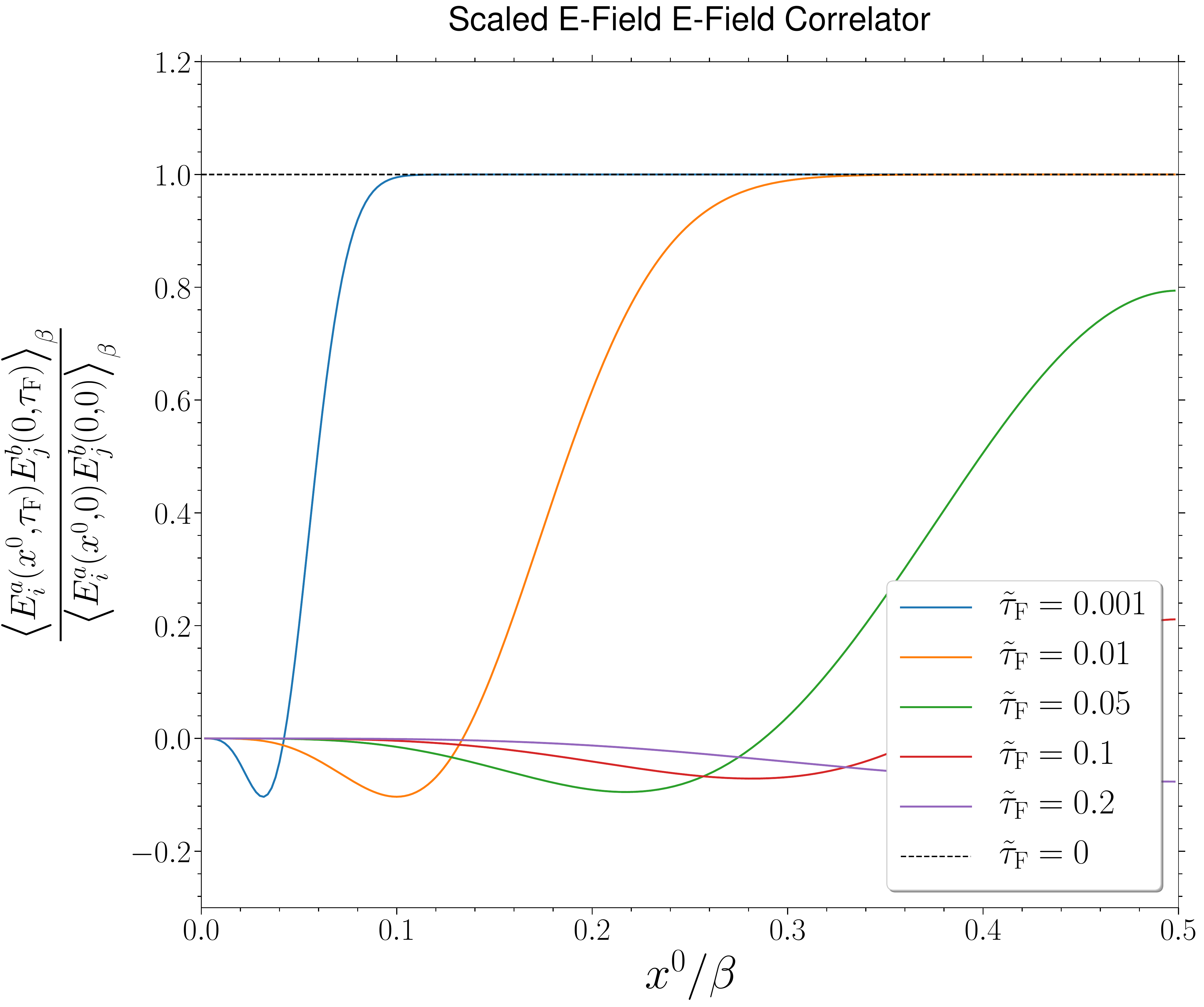}
 \caption{\label{fig: EE Sum Plot} Left: plot of the free-theory electric-field electric-field correlation, Right:the same, normalized to the unflowed behavior}
\end{figure}

The sum in \Eq{eq: EE Temp Sum} can be evaluated numerically. In
\figref{fig: EE Sum Plot}, the behavior of the correlator is shown for
different values of $\ttf$. The black curve is the analytic result for
zero flow from \Eq{eq: EE Temp Sum}. The blue curve related to a flow
time of $\ttf= 0.001$ is hidden under the zero flow curve for
$x^{0}/\beta > 0.2$. \figref{fig: EE Sum Plot} shows that as we
increase the amount of flow, the $x^0$ range for which the correlator
remains almost unchanged gets narrower; for the larger flow times
shown, the two never coincide.  Therefore the amount of flow which we
can ``get away with'' is $x^0$ dependent, which should not be too
surprising.

\begin{figure}[tb]
\centering
 \includegraphics[width = 0.8\textwidth]{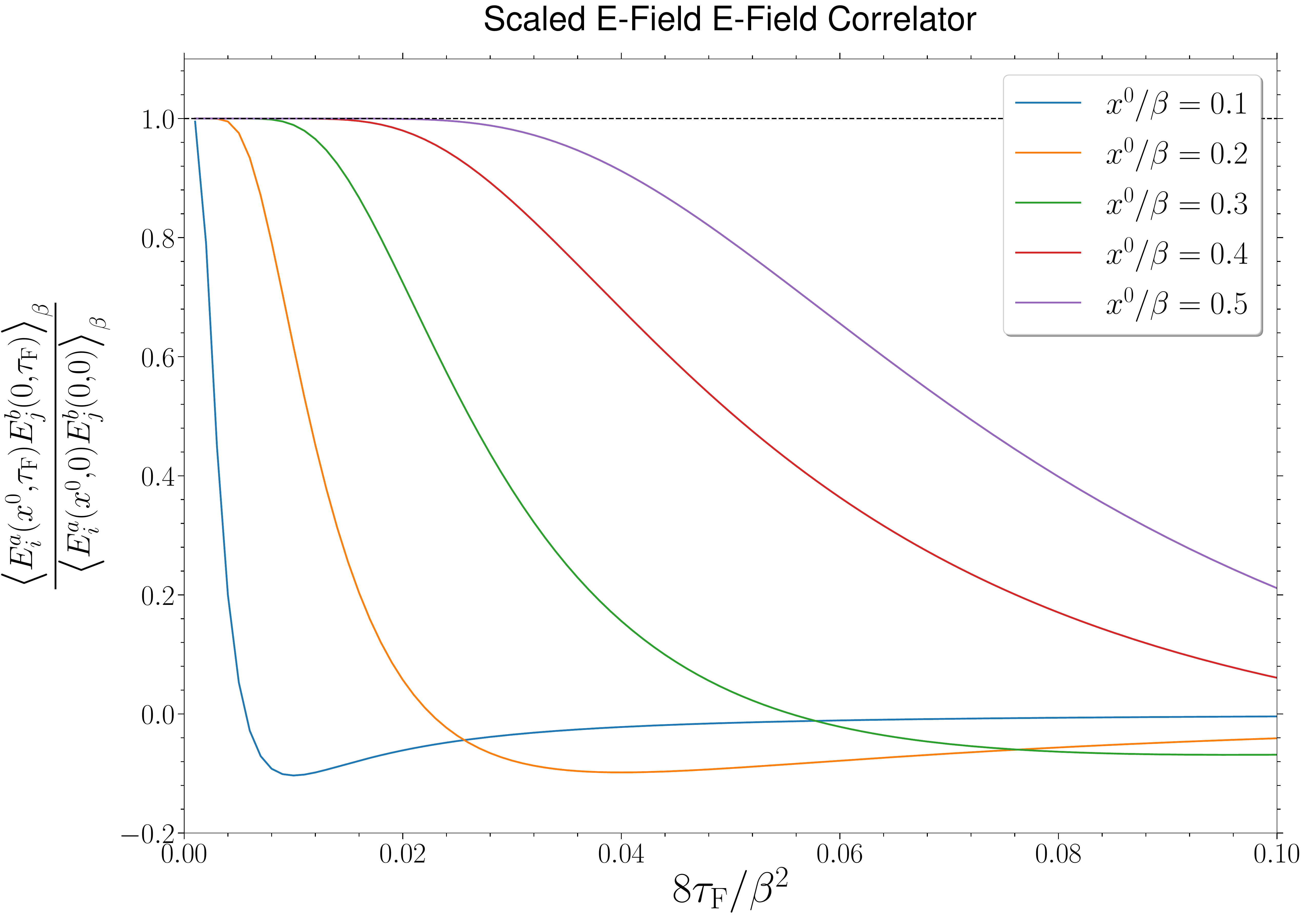}
 \caption{\label{fig: EE Plot x0 fix}
   Plot of the electric-field electric-field correlator normalized to
   the unflowed behavior at fixed
   time separations as a function of flow times.}
\end{figure}

It is also instructive to explore the correlator as a function of flow
time at fixed separation. \figref{fig: EE Plot x0 fix} shows a plot of
this function. The behavior of the function is exactly as
expected. For small flow times, the correlator shows a plateau of the
unflowed value and the amount of flow which damages the correlator
depends on the separation. If we use more flow, the correlator changes
sign.  For enough flow it becomes small as all fluctuations are damped
away.

\begin{table}[tb]
\begin{center}
\renewcommand{\arraystretch}{1.3}
\caption{List of the flow times needed to change the
  $\langle EE \rangle$  correlator by $1\%$ relative to the zero--flow
  value, for different $x^0$ separations.}
\begin{tabular}{|c|c|}
\hline 
$\frac{x^{0}}{\beta}$ & $\frac{8 \tf}{\beta^{2}}$ \\ 
\hline 
\hline
0.1 & 0.0011 \\ 
\hline 
0.2 & 0.0044 \\ 
\hline 
0.3 & 0.0099 \\ 
\hline 
0.4 & 0.0180 \\ 
\hline 
0.5 & 0.0274 \\ 
\hline 
\end{tabular} 
\end{center}
\label{tab: 1pro EE List}
\end{table}

In \tabref{tab: 1pro EE List} we show the maximum amount of flow
before the correlator changes by $1\%$ as a function of $x^0$.  We
believe that this can be used as a criterion for how much flow one can
``get away with'' in measuring the $EE$ correlator at a given $x^0$
value.

\subsection{Stress Tensor Two-Point Functions}
\label{subsec: TT}

The calculation of the stress tensor two-point correlator is similar
to the electric field correlator, except that each stress tensor
contains two field strengths.  Since there are two gauge field
propagators, there is now a double sum over images.
The connected stress tensor two-point function is
\beqa
&& \left\langle G_{\mu\sigma} G_{\nu\sigma}(x,\tf) \:
G_{\alpha\omega} G_{\beta\omega}(0,\tf) \right\rangle
\nn \\
& = &
\frac{\dA g^4}{16\pi^4} C_{\mu\nu\alpha\beta,abcdefgh}
\partial^x_a \partial^{x'}_b \partial^y_c \partial^{y'}_d  \\
& \times &
\left\{ \left[ \sum_n \frac{\delta_{eg}}{(x-y)_n^2}
    \left( 1 - e^{-\frac{(x-y)_n^2}{8\tf}} \right) \right]
   \left[ \sum_m \frac{\delta_{fh}}{(x'-y')_m^2}
       \left( 1 - e^{-\frac{(x'-y')_m^2}{8\tf}} \right) \right] \right.
     \nn \\
& & {} +    
     \left. \left.
      \left[ \sum_n \frac{\delta_{eh}}{(x-y')_n^2}
    \left( 1 - e^{-\frac{(x-y')_n^2}{8\tf}} \right) \right]
   \left[ \sum_m \frac{\delta_{fg}}{(x'-y)_m^2}
       \left( 1 - e^{-\frac{(x'-y)_m^2}{8\tf}} \right) \right]
     \right\} \right|_{x=x',y=y'=0} \,, \nn
\eeqa
where we have introduced the Lorentz structure
\be
\label{eq: TT lorentz}
\tensor{C}{_\mu_\nu_\alpha_\beta_,_a_b_c_d_e_f_g_h} =
\left( \tensor{\delta}{_\mu_a} \tensor{\delta}{_\sigma_e} {-}
\tensor{\delta}{_\mu_e} \tensor{\delta}{_\sigma_a} \right)
\left( \tensor{\delta}{_\nu_b} \tensor{\delta}{_\sigma_f} {-}
\tensor{\delta}{_\nu_f} \tensor{\delta}{_\sigma_b} \right)
\left( \tensor{\delta}{_\alpha_c} \tensor{\delta}{_\omega_g} {-}
\tensor{\delta}{_\alpha_g} \tensor{\delta}{_\omega_c} \right)
\left( \tensor{\delta}{_\beta_d} \tensor{\delta}{_\omega_h} {-}
\tensor{\delta}{_\beta_h} \tensor{\delta}{_\omega_d} \right).
\ee
The derivatives can be applied for each sum separately,
\be
\left. \partial^x_a \partial^y_c
\frac{\delta_{eg}}{(x-y)^2_n}
\left( 1 - e^{-\frac{(x-y)^2_n}{8\tf}} \right) \right|_{y=0}
= \frac{4 \delta_{eg}}{\pi^2 x_n^4}
\left( \delta_{ac} A_n(x,\tf) + \frac{x_{na} x_{nc}}{x_n^2}
B_n(x,\tf) \right)
\ee
with the dimensionless scalar functions defined as
\beqa
A_{n}(x,\tf) &=& \frac{1}{2}\left(1 - (1+\xit_n^2) e^{-\xit_n^2}
\right) ,
\\
B_{n}(x,\tf) &=& -2 + \Big( 2 + 2 \xit_n^2 + \xit_n^4 \Big)
e^{-\xit_n^2} \,.
\eeqa
Because at leading order $T_{00} = -T_{ii}$, there are three
independent stress-tensor correlators (at vanishing spatial momentum)
for which we can apply these formulae,
\be
\left\langle \tensor{T}{_0_0} \tensor{T}{_0_0} \right\rangle_
             {\beta},~~ \left\langle \tensor{T}{_0_i} \tensor{T}{_0_i}
             \right\rangle_ {\beta},~~ \left\langle \left(
             \tensor{T}{_i_j} - \thirdd \tensor{\delta}{_i_j}
             \tensor{T}{_k_k} \right)  \left( \tensor{T}{_i_j} -
             \thirdd \tensor{\delta}{_i_j}  \tensor{T}{_l_l} \right)
             \right\rangle \equiv
             \left\langle T_{ij}^{\mathrm{tr}} T_{ij}^{\mathrm{tr}}
             \right\rangle \, ,
\ee
which evaluate to
\beqa
\label{eq: T00T00 Temp }
\left\langle \tensor*{T}{_0_0}(x,\tf)
\tensor*{T}{_0_0}(0,\tf) \right\rangle & = &
\frac{\dA}{\pi^{4}} \sum_{n} \sum_{m}
\left[  12 \frac{A_{n} A_{m}}{x_{n}^{4} x_{m}^{4}}
  + 3 \frac{\left( A_{n} B_{m} + A_{m} B_{n}\right)}{x_{n}^{4}
    x_{m}^{4}}
  \right. \\ \nn && \left. \hspace{4em} {}
  + \left(  \left(x_{n} \cdot x_{m} \right)^{2}
  + \frac{1}{2} x_{n}^{2} x_{m}^{2}
  - 4 \vec{x}^{2} x_{n,0} x_{m,0} \right)
  \frac{B_{n} B_{m}}{x_{n}^{6} x_{m}^{6}} \right],
\\
\left\langle \tensor*{T}{_0_i}(x,\tf)
\tensor*{T}{_0_i}(0,\tf) \right\rangle & = &
\frac{\dA}{\pi^{4}} \sum_{n} \sum_{m}
\left[ 24 \frac{A_{n} A_{m}}{x_{n}^{4} x_{m}^{4}}
  + 6 \frac{\left( A_{n} B_{m} + A_{m} B_{n}\right)}
  {x_{n}^{4} x_{m}^{4}}
  \right. \\ \nn && \left. \hspace{4em} {}
  +  \left(  \left(x_{n} \cdot x_{m} \right)^{2}
  + 4 x_{n,0}^{2} \vec{x}^{2}  - x_{n,0}^{2} x_{m,0}^{2}
  + \vec{x}^{4}\right) \frac{B_{n} B_{m}}{x_{n}^{6} x_{m}^{6}} \right],
\\
\left\langle T_{ij}^{\mathrm{tr}}(x,\tf) 
T_{ij}^{\mathrm{tr}}(0,\tf)\right\rangle
 & = &
\frac{\dA}{\pi^{4}} \sum_{n} \sum_{m}
\left[ 80 \frac{A_{n} A_{m}}{x_{n}^{4} x_{m}^{4}}
  + 20 \frac{\left( A_{n} B_{m}
    + A_{m} B_{n}\right)}{x_{n}^{4} x_{m}^{4}}
  \right. \\ \nn && \hspace{4em} \left. {}
  +  \left( 10 \left(x_{n} \cdot x_{m} \right)^{2}
  - \frac{52}{3} \left(x_{n} \cdot x_{m} \right) \vec{x}^{2}
  \right. \right. \\ \nn && \hspace{6em} \left. \left. {}
  + \frac{10}{3} \left(x_{n}^{2} + x_{m}^{2} \right)
  \vec{x}^{2} + \frac{2}{3} \vec{x}^{4}\right)
  \frac{B_{n} B_{m}}{x_{n}^{6} x_{m}^{6}} \right].
\eeqa
These are the main analytic results of this section.

\begin{figure}[tb]
 \includegraphics[width = 0.48\textwidth]{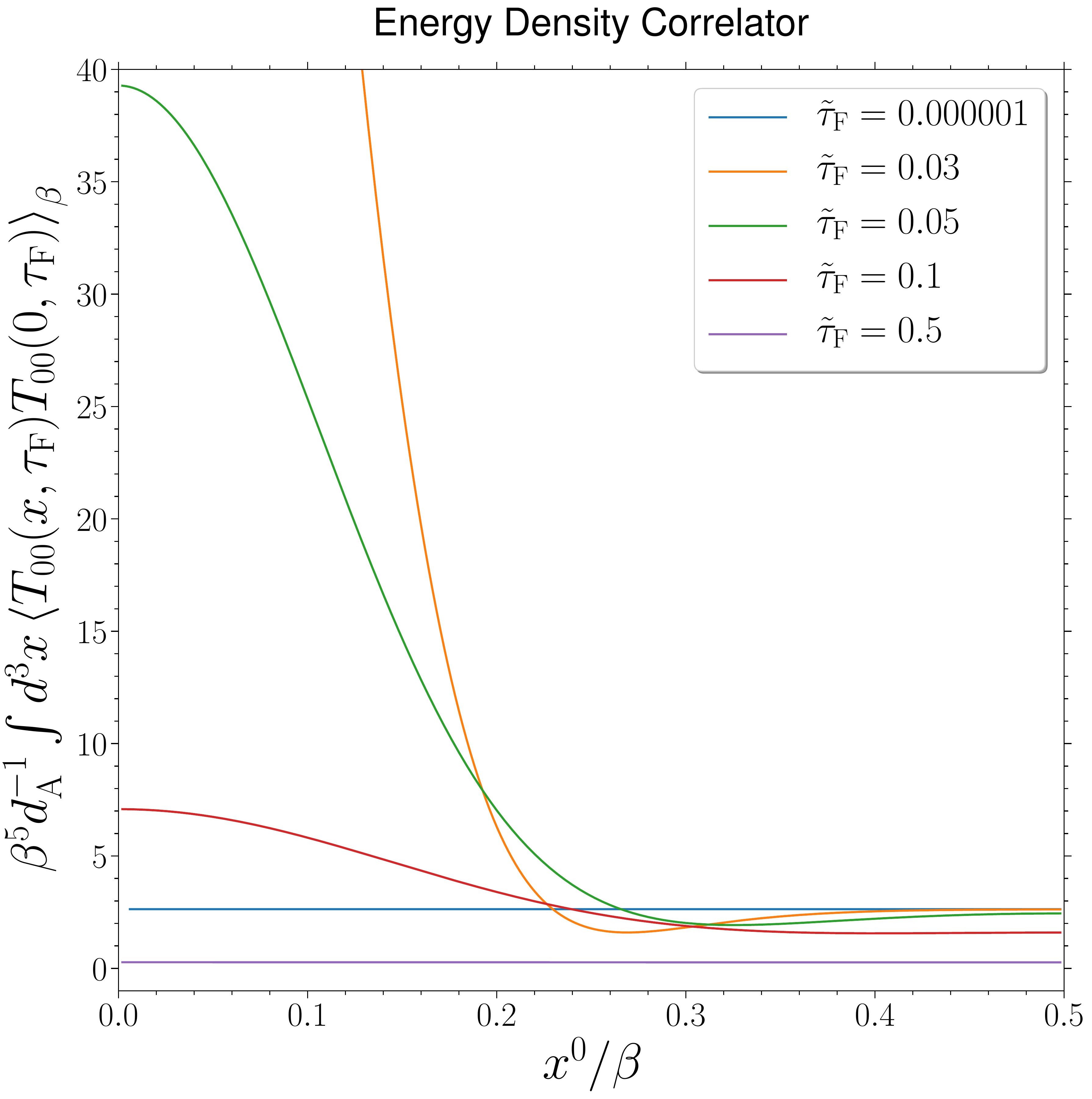} \hfill
 \includegraphics[width = 0.49\textwidth]{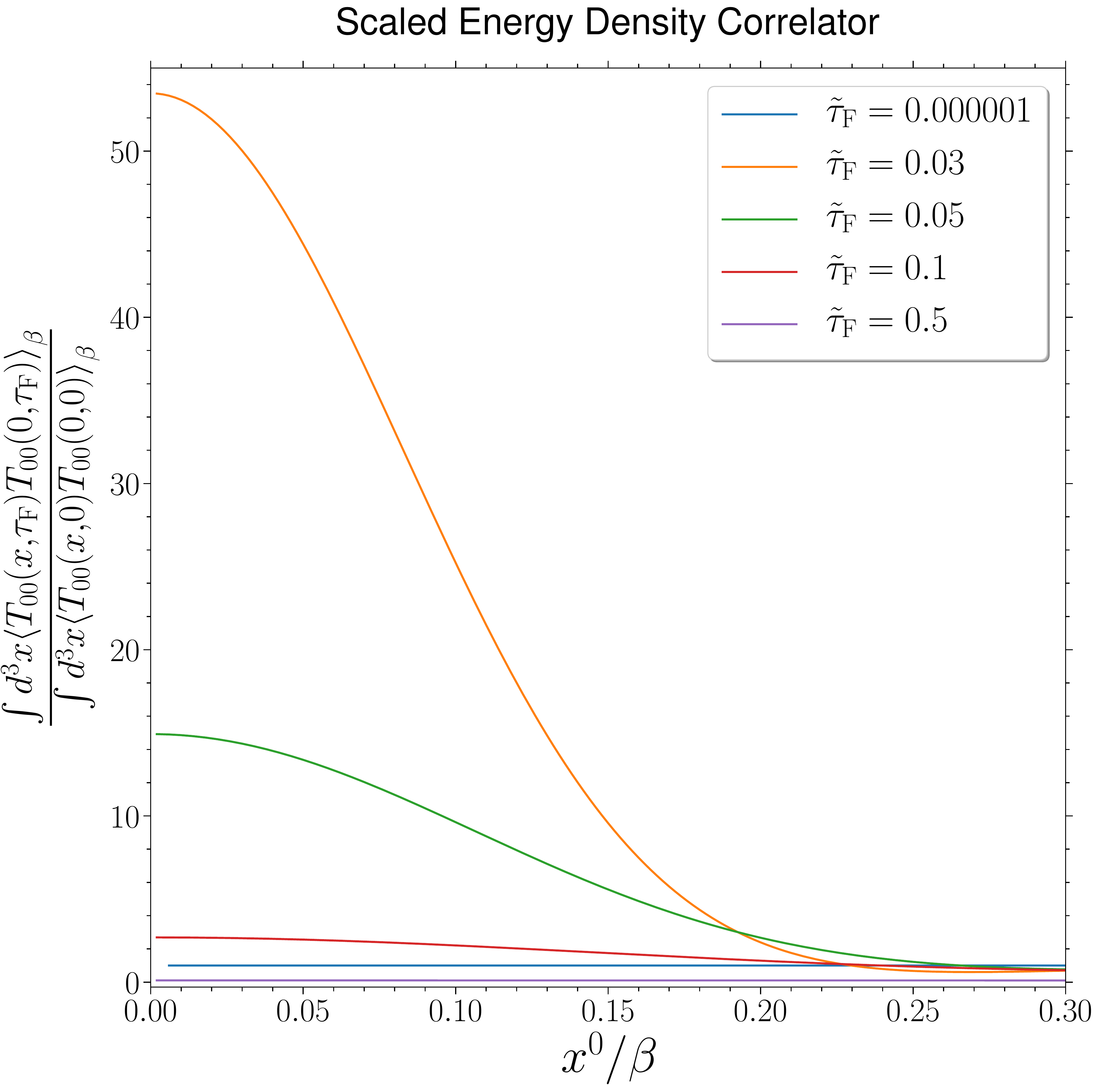}
 \caption{\label{fig: T00T00}
    Plot of the $\langle T_{00} T_{00} \rangle$ correlator at zero
    spatial momentum as a function of temporal separation, for
    selected values of flow time.}
\end{figure}

We are interested in the $\vec p=0$ channel and thus need
to integrate over $\int d^3 x$. At this point we resort to a
numerical evaluation.
In \figref{fig: T00T00} the results for the energy density--energy
density component
$\left\langle \tensor*{T}{_0_0}\tensor*{T}{_0_0} \right\rangle$
are shown.  Energy conservation implies that for vanishing flow time
the correlator should be a flat line at a value set by the heat
capacity, which in the free theory is
$\frac{4\pi^{2}\dA}{15\beta^5}$.  Such consequences of stress
conservation hold up to exponentially small corrections so long
as $\xit^2 \gg 1$.  However, for larger flow extents, operators
effectively overlap, and contact terms contaminate consequences of
stress conservation.  Therefore, when the flow depth approaches the
squared separation, the constancy of the
correlator will be lost.  This is indeed what we observe.  For large
values of flow time $\ttf \gg 1$, the correlator becomes flat again,
as it is dominated by the time-independent zero Matsubara frequency
contribution.

\begin{figure}[tb]
 \includegraphics[width = 0.49\textwidth]{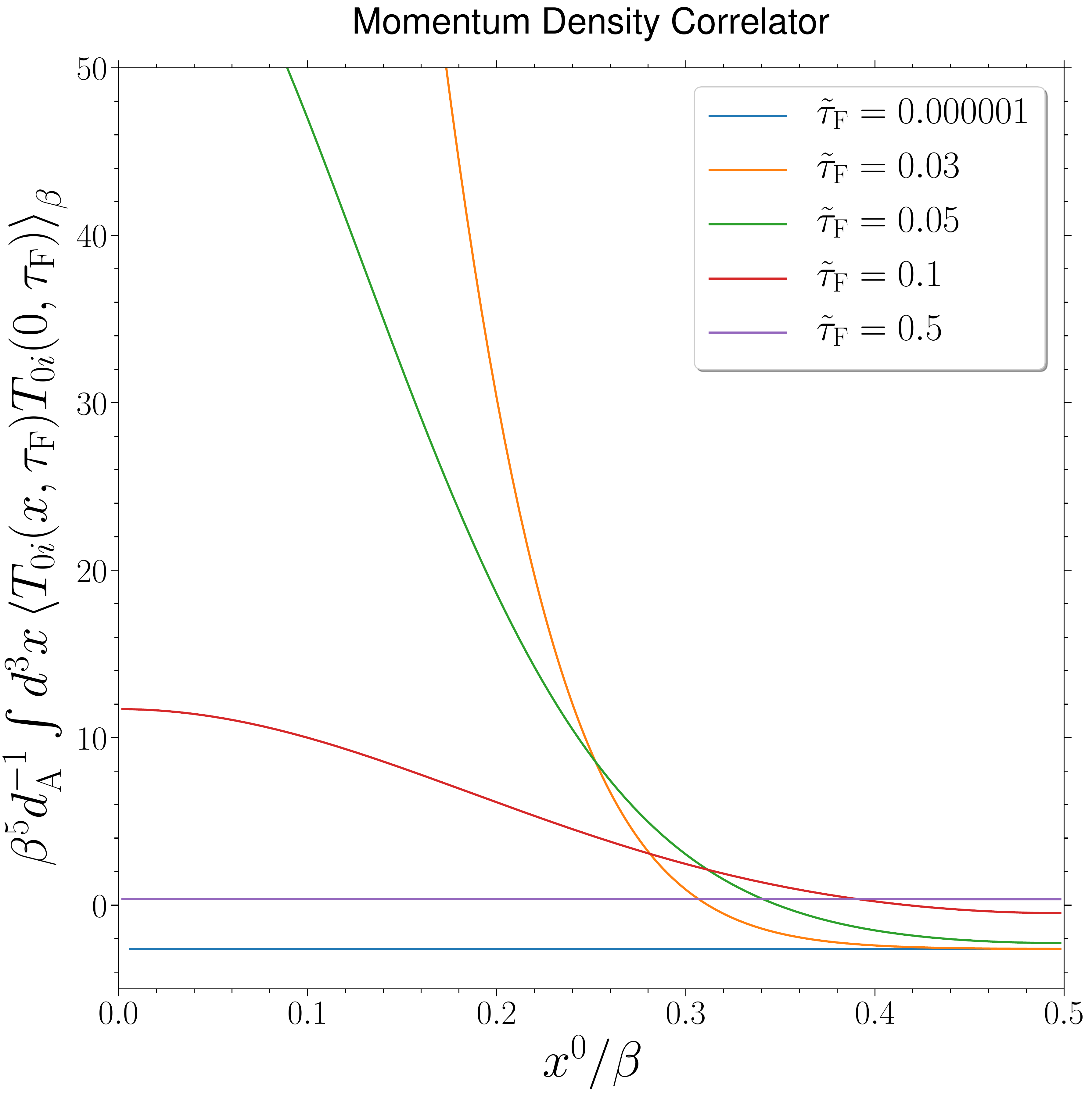} \hfill
 \includegraphics[width = 0.49\textwidth]{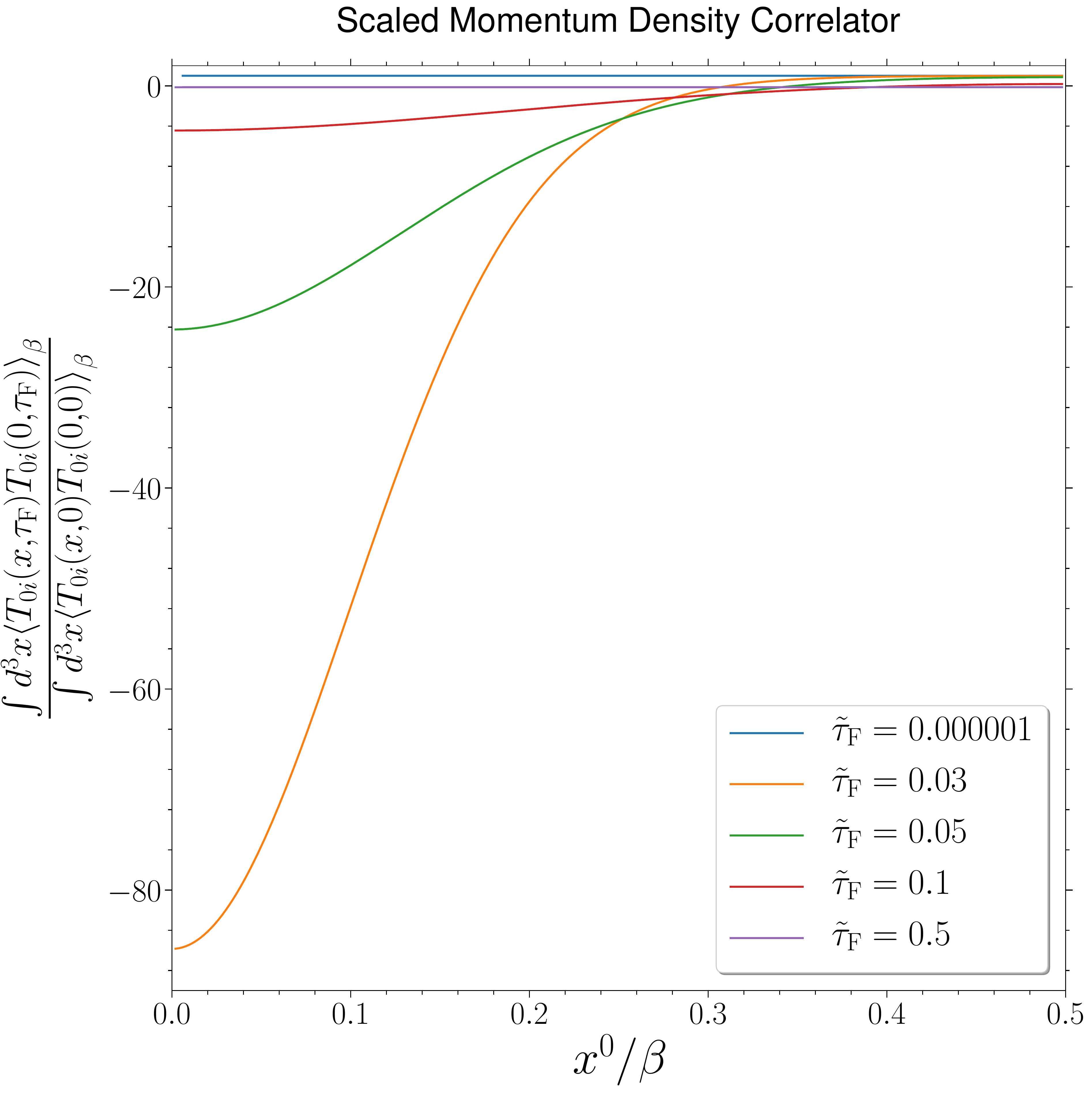}
 \caption{\label{fig: T0iT0i}
    Momentum density correlator as a function of temporal separation
    at selected flow depths.}
\end{figure}

The momentum--momentum component
$\left\langle \tensor{T}{_0_i} \tensor{T}{_0_i}\right\rangle$
is related to momentum fluctuations in the medium.  Without flow,
it should be constant and negative, with value set by the enthalpy
density times temperature,
$\frac{4 \pi^{2}\dA}{15\beta^5}$.
The behavior
under flow is shown in \figref{fig: T0iT0i}.  The flow time dependence
is similar to that for the $\langle T_{00} T_{00} \rangle$ correlator,
for the same physical reasons.

\begin{figure}[htb]
\centering
 \includegraphics[width = 0.493\textwidth]{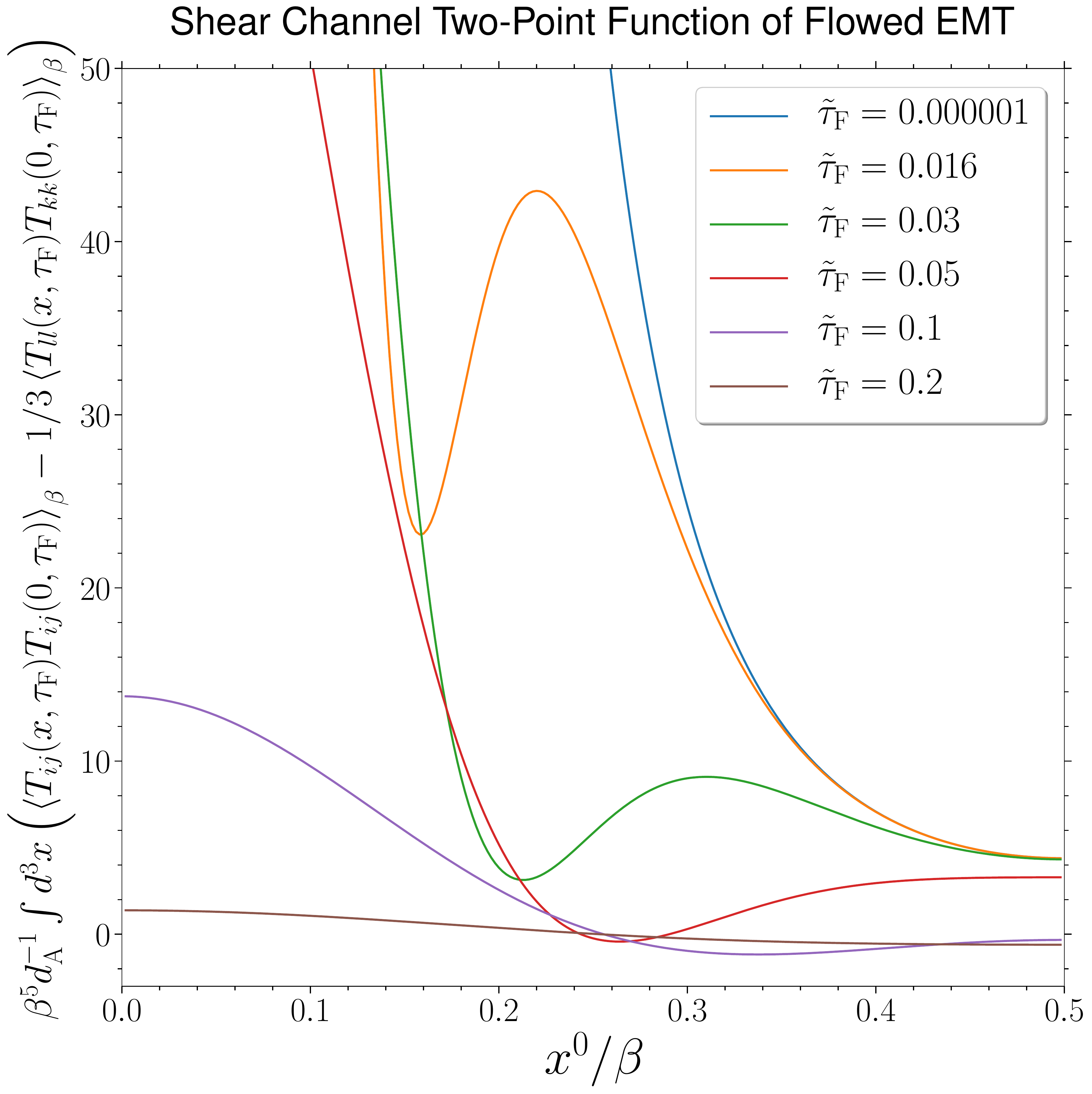}
 \includegraphics[width = 0.49\textwidth]{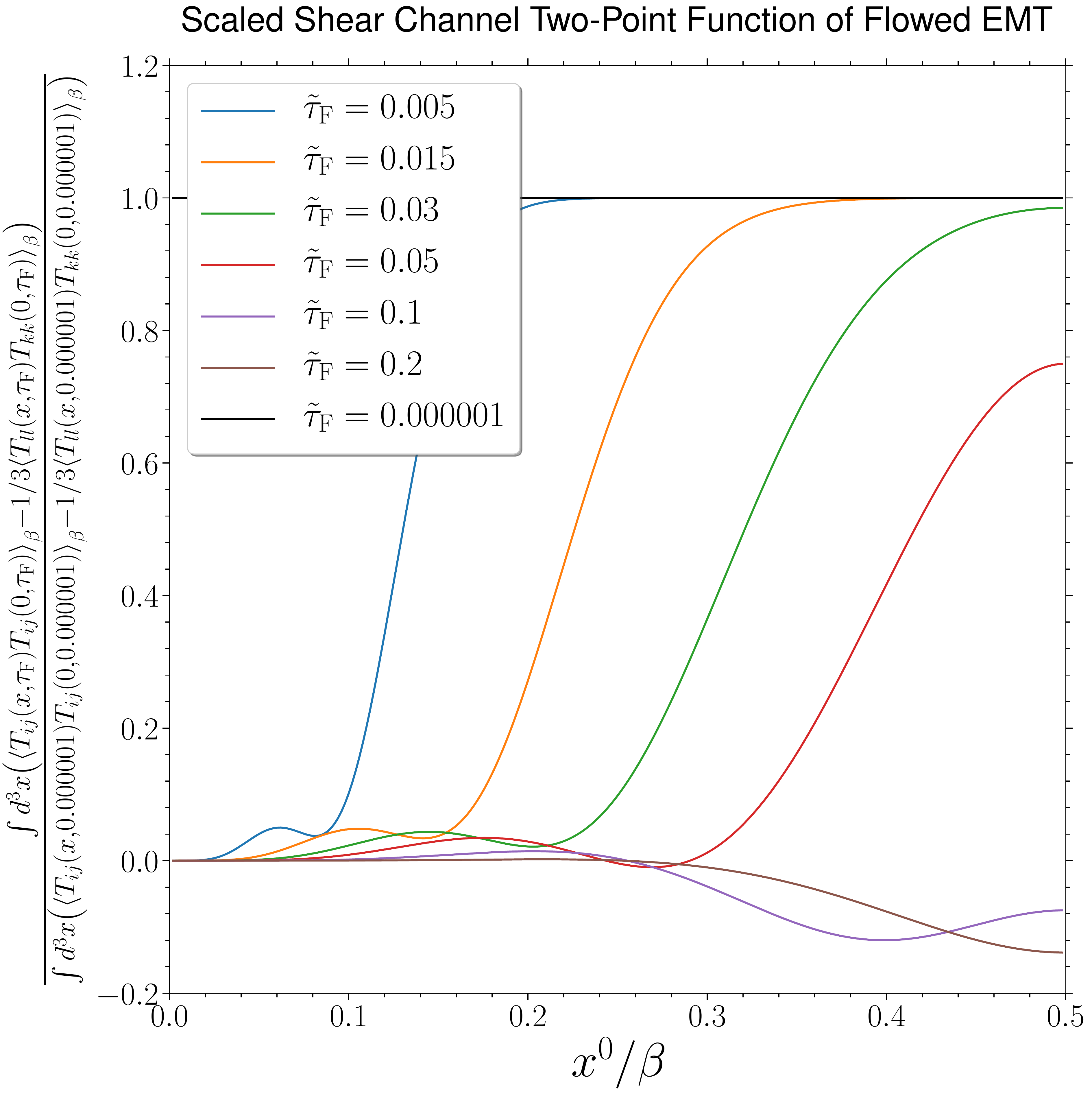}
 \caption{\label{fig: TijTij}
  Left:  plot of the shear channel two--point function of the
  energy--momentum tensor as a function of temporal separation.
  Right:  the same but normalized to the unflowed result.}
\end{figure}

The stress--stress component $\left\langle \left(\tensor{T}{_i_j} -
\thirdd \tensor{\delta}{_i_j} \tensor{T}{_l_l} \right)
\left(\tensor{T}{_i_j} - \thirdd \tensor{\delta}{_i_j} \tensor{T}{_k_k}
\right)\right\rangle$ is physically interesting because its
continuation to a spectral function determines the shear viscosity
\cite{Zubarev,Karsch:1986cq,Meyer:2007ic,Meyer:2011gj}.
Because it is not constrained by
conservation laws, no short-distance cancellations occur and it shows
strong short-distance divergent behavior; the unflowed behavior is dominated
by the vacuum contribution which diverges at the origin. If we use
gradient flow, the correlator is finite at the origin and for
intermediate flow times $0.01 < \ttf <0.1$ we find a non--trivial
behavior. The numerical results are presented in
\figref{fig: TijTij}.  For large flow times the zero
Matsubara frequency again dominates the correlator, which is nearly
$x^0$ independent.

The main result of this numerical evaluation is that if we are
using flow to suppress fluctuations in our correlators, then the
$\langle T_{00} T_{00}\rangle$ and $\langle T_{0i} T_{0i}\rangle$
correlators are best evaluated at $x^0=\beta/2$ and with at most
$8\tf/\beta^2 < 0.027$.  For $\langle T_{ij} T_{ij} \rangle$ one
should use the same $\tf$ values as for the electric field correlator
with the same $x^0$ value.

\section{Discussion and conclusions}
\label{discussion}

Gradient flow successfully reduces short-distance fluctuations, which
is a boon for reducing statistical fluctuations in the lattice
determination of local operator correlation functions.  Therefore
there is an interest in applying it to lattice measurements of
correlation functions.  Here we made a first exploration of how
reliable this approach may be at finite temperature, for the
evaluation of the energy density $T_{00}$ and of electric field and
stress tensor two-point functions.  At lowest order in perturbation
theory, we found that the energy density of the thermal bath is
obtained reliably provided that the flow depth obeys
$\tf < 0.015 \beta^2$ (or $\tf/a^2 = 0.015 N_t^2$ on the lattice),
whereas a 2-point function of field strengths or stress tensors
separated by a distance $x^0$ is reproduced reliably for
$\tf < 0.014 (x^0)^2$ (or $\tf/a^2 = 0.014 (\Delta N_t)^2$ on the
lattice, where $\Delta N_t$ is the minimum number of lattice units of
separation between the two operators to be evaluated).  Exceeding this
amount of flow causes
contact-term contamination in the correlator, either between operators
or between an operator and its periodic images.  However, below this
amount of flow, the effect of flow on the correlation function due to
these effects is exponentially small, and consequences of symmetries
such as stress tensor conservation are preserved up to exponential
corrections.

It would be valuable to extend this study to the loop level, to see
how operator renormalization, the Wilson line appearing in the
definition of the electric field two-point function, and other
interaction effects enter, and to check whether these effects modify
our conclusions.

\section*{Acknowledgments}

We thank the Technische Universit\"at Darmstadt and its Institut f\"ur
Kernphysik, where this work was conducted.
This work was supported by the Deutsche Forschungsgemeinschaft (DFG)
through the grant CRC-TR 211 ``Strong-interaction matter under extreme
conditions.''

\bibliographystyle{unsrt}
\bibliography{refs}

\end{document}